\documentclass{article}
\newcommand{\bs}{\boldsymbol}
\newcommand{\mc}{\mathcal}
\newcommand{\mb}{\mathbf}
\newcommand{\mbb}{\mathbb}

\usepackage{amsmath,amssymb,amsfonts}
\usepackage{spconf,amsmath,graphicx}
\usepackage{xcolor}
\usepackage{enumitem}
\setlist{nosep, leftmargin=14pt}

\usepackage{mwe} 

\usepackage{url} 

\title{Accelerating quantitative MRI using subspace multiscale energy model (SS-MUSE)}
\name{Yan Chen$^*$, Jyothi Rikhab Chand$^*$, Steven R. Kecskemeti$^\dag$, James H. Holmes$^\ddag$, Mathews Jacob*\thanks{This work is supported by NIH grants R01AG067078, R01AG087159, R01HD108868, P50HD103556 and R01EB019961.}}
\address{$^*$University of Virginia, $^\dag$ University of Wisconsin-Madison, $^\ddag$ University of Iowa }
\begin{document}
%
\maketitle
\begin{abstract}
Multi-contrast MRI methods acquire multiple images with different contrast weightings, which are used for the differentiation of the tissue types or quantitative mapping. However, the scan time needed to acquire multiple contrasts is prohibitively long for 3D acquisition schemes, which can offer isotropic image resolution. While deep learning-based methods have been extensively used to accelerate 2D and 2D + time problems, the high memory demand, computation time, and need for large training data sets make them challenging for large-scale volumes. To address these challenges, we generalize the plug-and-play multi-scale energy-based model (MuSE) to a regularized subspace recovery setting, where we jointly regularize the 3D multi-contrast spatial factors in a subspace formulation. The explicit energy-based formulation allows us to use variable splitting optimization methods for computationally efficient recovery. 

\end{abstract}

\section{Introduction}
\begin{figure*}[ht!]
    \centering   
    \includegraphics[height=.25\textheight]{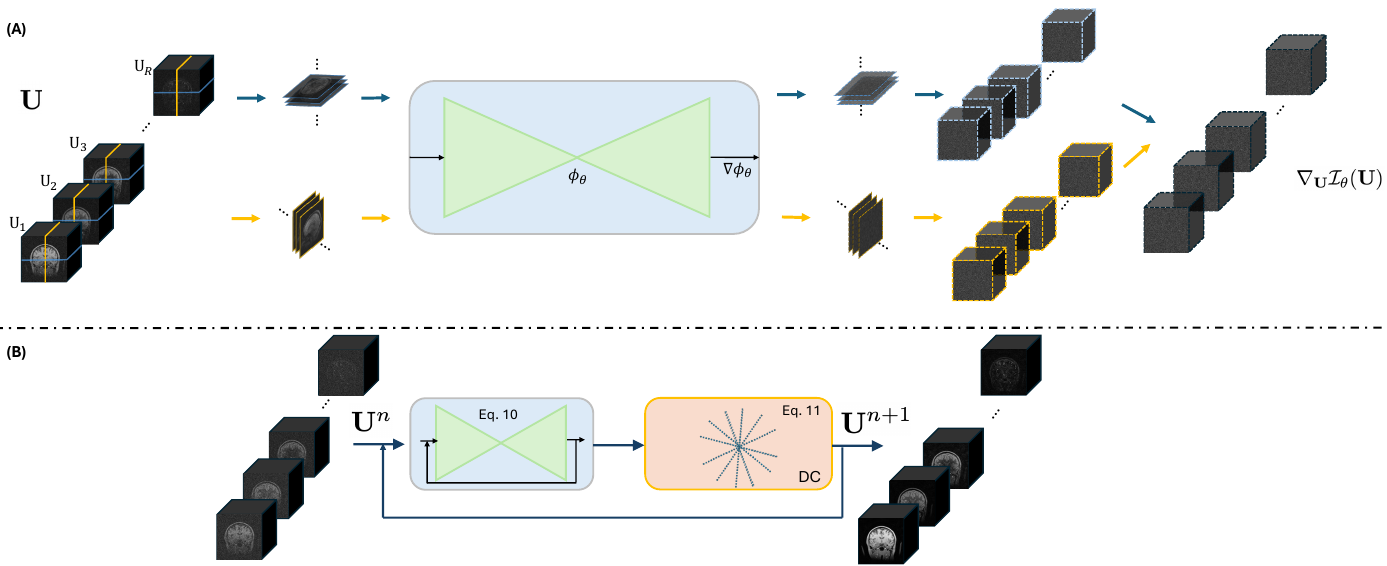}
       \caption{Method overview: (A) Evaluation of the energy $\mc I_{\theta}(\mb U)$ of 4D spatial factor and the associated score using a 2D energy model $\phi_{\theta}$. The score of the spatial factor $\mb U$ is the gradient of the energy $\mc I_{\theta}(\mb U)$ with respect to $\mb U$. We define the energy of the 4D $\mathbf U$ as the sum of the energies of the 2D slices in the $x-z$ and $y-z$ planes, respectively. This translates to the score of $\mc I_{\theta}(\mb U)$ being the concatenation of the scores of the respective slices in the two orientations, followed by an average of two directions. (B) Maximum a-posteriori estimation of the data given $k$-space measurements using variable splitting: The proposed variable splitting strategy results in an iterative algorithm that alternates between Eq. \eqref{eq:subProb}, which is an image domain denoising of the auxiliary variable $\bs{\mc Z}$, and the data consistency (DC) step specified by Eq. \eqref{dcsubproblem}. Once the spatial basis functions are derived, the source images at different inversion times can be generated using Eq. \eqref{eq:lowrank}. }
    \label{fig:mainfig}
    \vspace{-0.15in}
\end{figure*}
\label{sec:intro}

Relaxometry-based MRI techniques such as MR fingerprinting and MPnRAGE \cite{kecskemeti2016mpnrage} capture a series of data points across the temporal evolution of magnetization, which are then used to estimate key relaxation parameters. 
For example, MPnRAGE uses a series of inversion pulses interleaved with 3-D radial gradient echo readouts with pseudo-random angular view ordering within each inversion period. The data is used to recover hundreds of 3D volumes during longitudinal magnetization relaxation, which enables multi-contrast imaging and quantitative $T_1$ mapping. The ability to provide qualitative information typically collected using weighted acquisitions such as MP-RAGE \cite{mugler1990three}, while offering quantitative parameters such as $T_1$, makes this approach attractive. However, the relatively long acquisition time of this approach restricts its clinical adoption.

Low-rank/subpace approaches have been widely explored for dynamic MRI and multi-contrast imaging to reconstruct time-resolved images from undersampled measurements. These methods exploit the spatial and temporal correlation in the image time-series by explicitly factorizing the image series into spatial and temporal coefficients \cite{ravishankarLowRankAdaptiveSparse2017,zhaoAcceleratedMRParameter2015}. This approach also reduces the computational and memory burden by eliminating the need to work with the full image time-series \cite{zhaoAcceleratedMRParameter2015}. The temporal coefficients are typically derived from the auxiliary data obtained from Bloch equation simulation. The spatial coefficients are estimated from the data, with a sparsity prior in the transformed domain to resolve the aliasing artifacts in undersampled settings. Typical transform operators include wavelet transform and finite differences, applied along either spatial or temporal dimensions.

In recent years, deep-learned regularizers have emerged as powerful alternatives to classical regularizers for inverse problems\cite{qiuDirectSynthesisMulticontrast2023}. For instance, the model-based deep learning (MoDL) approach uses a CNN regularization module in a reconstruction algorithm \cite{aggarwal2018modl}; the weights of the CNN module were learned in an end-to-end fashion by unrolling the algorithm. Several authors have extended the unrolled optimization scheme with subspace models using either supervised or self-supervised learning strategies \cite{sandino2021deep,chen2022data,jun2024zero,aggarwal2018modl,liuMRParameterMapping2021}. Although E2E training offers good performance, the unrolling strategy is challenging for the high-resolution 3D + time setting in this work due to the memory demand. In addition, extensive training data is not available 
A self-supervised Deep Factor Model (DFM) has been recently introduced to accelerate MPnRAGE with reduced memory demand compared to E2E by training a CNN in a self-supervised manner\cite{chen2023deep}. However, a key challenge with self-supervised methods like DFM is the long run-time, because training involves learning the parameters of the neural network based on the k-space error, evaluated using the computationally expensive 4D NUFFT.  

We introduce a subspace-based method to jointly recover 3D multi-contrast images, where we use the multi-scale energy (MuSE) regularization \cite{chand2024multi} on the spatial coefficients in a plug-and-play (PnP) fashion. The main distinction from \cite{chand2024multi} are:
\begin{itemize}
    \item We apply MuSE as a prior on the spatial factors in a subspace recovery setting. Unlike classical PnP methods \cite{venkatakrishnan2013plug,ahmad2020plug}, the CNN in MuSE need not be constrained as a contraction to guarantee convergence. This relaxation translates to superior denoising performance and consequently improved performance \cite{chand2024multi}. While the performance is comparable to E2E deep learning models, the PnP energy model is significantly more memory efficient than the unrolled E2E model. In addition, the PnP approach results in models that are applicable to arbitary undersampling rate, unlike E2E methods that are trained for specific settings. 
    \item We introduce a variable splitting scheme with fast convergence to minimize the negative log posterior, which is explicitly defined unlike PnP schemes.
    \item We extend the 2D MuSE model to the 4D setting by considering the 4D energy as the sum of the 2D MuSE energies of the individual slices. This approach enables us to train the model from the limited training datasets. This approach is the extension of the 3D approach in \cite{lee2023improving} to the 4D setting with contrast changes.

\end{itemize}

\section{Methods}
\vspace{-0.05in}

\label{sec:method}

\subsection{MPnRAGE sequence}
\vspace{-0.1in}
\label{ssec:dataAcquisition}
We used a 3D radial inversion recovery sequence MPnRAGE which consists of an adiabatic inversion pulse followed by a train of 385 radial gradient echoes, and a delay of 503.5 ms for magnetization recovery. We acquire the data with FOV = $256\times 256\times 256$ mm$^3$ and an isotropic resolution of $1$ mm$^3$, TR = $4.88$ ms. A variable flip angle approach, where the first 304 gradient echoes were acquired with a $4^{\circ}$ flip angle and the last 81 echoes with $8^{\circ}$, was used.

    \vspace{-0.15in}
    \subsection{Forward model and subspace reconstruction}
    \vspace{-0.1in}
    
The image volume $\bs X \in \mathbb{C}^{M\times T} $ consists of $T$ $T_1$-weighted images, each with $M$ voxels.   
    \begin{equation}
     \vspace{-0.05in}
      \bs  X = [  
                 \bs x_{1},  \bs x_{2}, \bs x_{3}, \bs \ldots, \bs x_{T} 
                ]
    \end{equation}
    We represent the signal at a particular time as:
    \begin{equation} 
        \vspace{-0.05in}
    \label{static} \mb b_ {\tau} = \mc A_{\tau} \bs x_{\tau} + \mb n_{\tau},\quad \mb n_{\tau}  \sim \mc N(0,\sigma^2 \mb {I})
    \vspace{-0.05in}
    \end{equation}
    Here, $\mathcal A_{\tau}$ denote the forward model for multichannel Fourier measurements and the $k$-space trajectory at ${\tau}$ for each excitation as $\mc A_{\tau}$ and $\mb b_{\tau}$ respectively,  $\mb n_{\tau}$ denotes the Gaussian noise in $k$-space at TI $\tau$.
    The subspace approach models the image series as below:
    \begin{equation}
        \vspace{-0.05in}
        \label{eq:lowrank}
        \mb X = \mb U \mb V
    \end{equation}
    where $\mb U\in \mathbb{C}^{M\times R}$ and $\mb V \in \mathbb{C}^{R\times T}$ contain $R$ basis functions ($R\ll T$), respectively. For 3D images, we define the matrix size $M =M_x\times M_y\times M_z$ in 3 dimensions.  The temporal factor $\mb V$ is pre-determined from simulated signals using principal component analysis (PCA). Recovering $\mb X$ translates to the reconstruction of spatial factor $\mb U$ with a greatly reduced degree of freedom. 
    \vspace{-0.15in}
    \subsection{Multiscale energy model (MuSE)}
    \vspace{-0.1in}
    
    MuSE models the negative logarithmic prior distribution of 2D images using a CNN. We generalize this approach to the multidimensional setting by modeling the prior distribution of the spatial factors as:
    \begin{equation}
         p_{\theta}(\mb U) = \frac{1}{Z_{\theta}}\exp{\bigg(\frac{-\mc I_{\theta}(\mb U)}{\sigma^2}\bigg)}.
    \end{equation}
    Here, $Z_{\theta}$ is a normalization constant to ensure the integration of $p_{\theta}(\mb U)$ is one, $\theta$ denotes the parameters of energy model $\mc I_{\theta}(\mb U)$. Training 3D/4D CNN-based energy model for $\mb U$ would require large amounts of fully-sampled data in 4D, which is unfortunately not available in our setting. In addition, such models will have a large memory footprint, which is also a challenge. Motivated by \cite{lee2023improving}, we propose to use a 2D energy model. In this work, we use the 2D energy model $\phi: \mathbb C^{M_y\times M_z} \rightarrow R+$ in \cite{chand2024multi}:  
     \begin{equation}
        \phi_{\theta}(\bs u) = \dfrac{1}{2}\|\bs u- \psi_{\theta}(\bs u)\|_2^2,
    \end{equation}
    where $\psi_{\theta}(\cdot): \mbb C^{M_y\times M_z} \rightarrow \mbb C^{{M_y\times M_z}}$ is a DRUnet CNN, and $\bs u$ denotes a 2D image slice.  We learn the parameters of $\psi$ using multi-scale denoising score matching:
    \begin{equation}
        \theta^* = \arg \min_{\theta} \mbb E_{\sigma} \big( \mbb E_{\bs u} \mbb E_{\bs z}\big [\gamma(\sigma)\|\nabla_{\bs u} \mc \phi_{\theta} (\bs u + \sigma \bs z) - \sigma \bs z \|_2^2\big ]  \big),
    \end{equation}
    Note that we train the model with noise at  different scales, which makes the network robust to local minima issues. n the above equation, $\bs z\sim \mc N(0,\mb I)$, and $\gamma(\sigma)$ is a positive weighting function depending on variance $\sigma$. The multiscale training also results in an energy, which measures the distance of the data point from the data manifold.
    
    We express the energy of the 4D volume $\mc I_{\theta}(\mathbf U) = \frac{1}{2}\big (\mc I^x_{\theta}(\mb U)+\mc I^y_{\theta}(\mb U)\big )$. Here, $\mc I^x_{\theta}(\mb U)$ is the sum of the 2D energies of the slices of the basis images, extracted along the $y-z$ orientation:
    \vspace{-1em}
      \begin{equation}
        \mc I^x_{\theta}(\mb U) = \sum_{r=1}^R \sum_{m_x=1}^{M_x} \phi_{\theta}(\mc B_{m_x,r} \mb U)
    \end{equation}
    Here, $\mc B_{m_x,r}$ extracts the $r_{th}$ basis function at slice index $m_x$ in $y-z$-plane. We note that $\sum_{r=1}^{R}\sum_{m_x=1}^{M_x}\mc B_{m_x,r}^{H}B_{m_x,r}\mb U = \mb U$. We also define $I^y_{\theta}(\mb U)$ in the same fashion by slicing along the $x-z$ plane.

    
    \vspace{-0.15in}
    \subsection{Maximum a-posteriori (MAP) estimation using MuSE}
    \vspace{-0.1in}
    We model the prior distribution of $\mb U$ using MuSE and pose the objective function of MAP given measurements $\bs b$ as follows:
    \begin{equation}
    \begin{split}
        \mb U^* 
        &=\arg \min_{\mb U} {\dfrac{1}{2}}\|\mc A(\mb U \mb V)- \mb b\|_2^2 + \lambda \mc I_{\theta}(\mb U)
    \end{split}
    \label{eq:map}
    \end{equation}
    where $\bs b = [\bs b_{1}, \bs b_{2},...,\bs b_{T}]$ and $\mc A =[\mc A_{1}, \mc A_{2},...,\mc A_{T}]$. 
    Solving \eqref{eq:map} is time-consuming due to the high computational complexity in evaluating $\mc A$ and $\mc A^H$, which are implemented using multichannel 3D non-uniform fast Fourier transforms (NUFFT). 
    To accelerate the convergence, we propose to solve Eq. \ref{eq:map} using variable splitting: 
    \begin{equation}
        \mb U^* =\arg \min_{\mb U, {\bs{\mc Z}}} \dfrac{1}{2}\|\mc A(\mb U \mb V)- \mb b\|_2^2 + \beta \|\mb U - \bs{\mc Z}\|^2 + \lambda \mc I_{\theta}(\bs{\mc Z})
        \label{eq:splitting}
    \end{equation}
    here, $\bs {\mc Z}\in \mathbb{C}^{M\times R}$ is an auxiliary variable. We consider $\mc I_{\theta}$ from two directions, $\mc I_{\theta}^x$ and $\mc I_{\theta}^y$. Hence, we can optimize Eq. \ref{eq:splitting} by alternating between two subproblems:
      \begin{eqnarray}\nonumber
      \label{eq:subProb}
            \bs{\mc Z}^{n+1} &=& \arg \min_{\bs{\mc Z}} \|\mb U^n - \bs{\mc Z}\|_2^2 + \frac{\lambda}{2\beta}\big ( \mc I_{\theta}^x({\bs{\mc Z}})+ \mc I_{\theta}^y({\bs{\mc Z}})\big)\\\\\nonumber
                    \label{dcsubproblem}
            \mb U^{n+1} &=& \arg \min_{\mb U} \dfrac{1}{2}\|\mc A (\mb U \mb V) - \bs b\|_2^2 + \beta \|\mb U - \bs{\mc Z}^{n+1}\|_2^2\\\vspace{-1em}
    \end{eqnarray}    
We solve \eqref{dcsubproblem} using conjugate gradients (CG) algorithm. We note that \eqref{eq:subProb} is a proximal map, which is evaluated using steepest descent. 
    
\vspace{-0.05in}

\section{Implementation details}
\vspace{-0.05in}

\subsection{Dataset}

Five healthy volunteers (2 female, 3 male, 20-50 years of age) were scanned on a clinical 3T scanner using a body transmit and 48-channel receive RF head coil. The 48 receive coils were compressed to 4 virtual coils using PCA. We estimated coil sensitivity maps using JSENSE. We acquired 224 inversion blocks in 9 minutes of scan time. In this work, we extracted 112 inversion blocks to test the methods, which corresponds to a 4.5-min scan time. 

\label{sec:dataprocessing}
We reconstructed the five volunteer scans using the self-supervised method, Deep Factor Model (DFM) \cite{chen2023deep} from the 9-min MPnRAGE scan. We split the data from the five volunteers into four for training and one for testing. We retrieve slices from all three axes. Each 2D slice was normalized such that the magnitude was less than one. Slices with low signal-to-noise ratio (SNR) were disregarded. This process yielded 930 complex 2D slices for training.
\vspace{-0.15in}
\subsection{Architecture and training of the MuSE model}
\vspace{-0.1in}
In this work, we consider image recovery with a rank of eight. i.e, we assume that $\mb U\in \mbb C^{256\times256\times256\times8}$. Each complex 2D slice is considered a two-channel real image. We implemented $\psi_{\theta} (\cdot)$ as a DRUnet with channels 64, 128, 256, and 512 using PyTorch. 
The standard deviations were uniformly chosen from 0 to 0.2. $\theta$ was trained for 80 epochs using the ADAM optimizer with a learning rate of $1\times 10^{-4}$ and batch size of 32. The weighting function $\gamma(\sigma)$ was chosen as $\frac{1}{\sigma}$ to balance a series of variances. 

\begin{figure}[t!]
    \centering
    \includegraphics[width=0.5\textwidth]{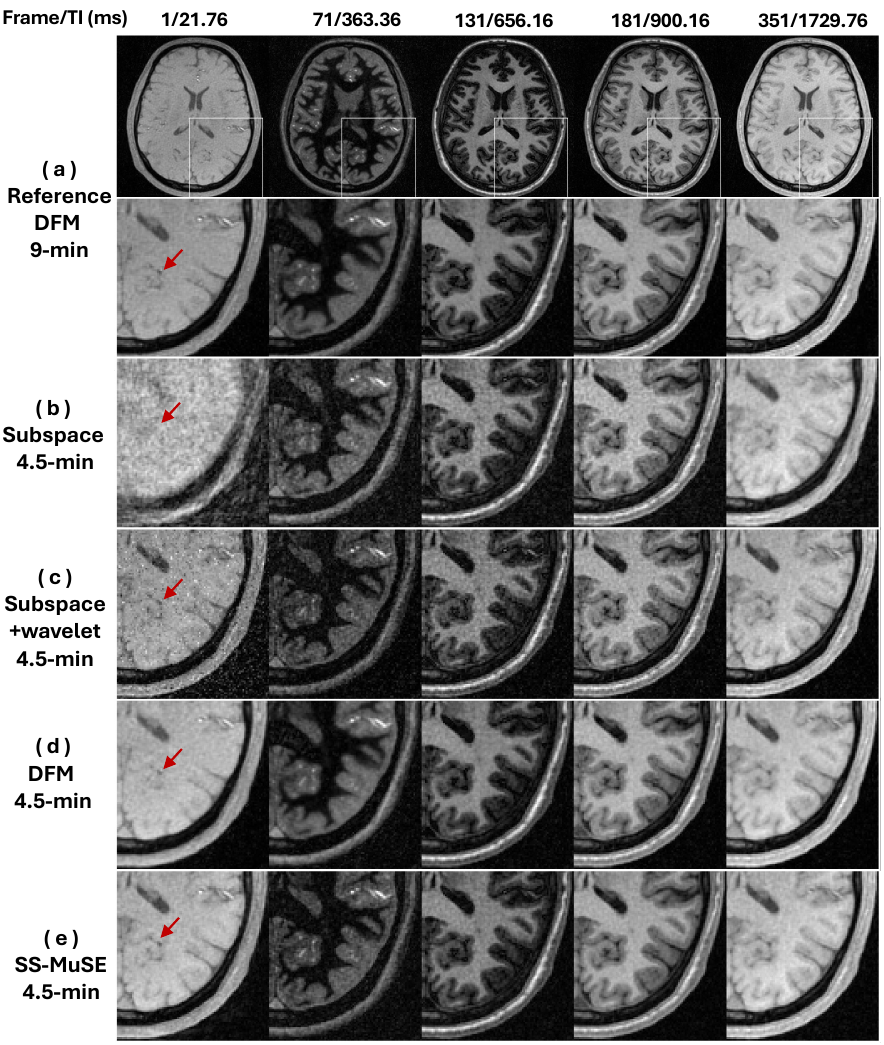}
    \caption{Comparisons of multi-contrast images from accelerated MPnRAGE scans: Five columns show multi-contrast images from five TIs with the frame index out of 384 and TI. (a): Reference images from a 9-min scan using DFM; The later rows show images from a 4.5-min scan using (b) Subspace; (c) Subspace + wavelet ; (d) DFM; and (e) SS-MuSE. SS-MuSE reduces aliasing artifacts compared with Subspace and Subspace + Wavelet methods while maintaining sharper edges than DFM for a 4.5-min scan.}
    \label{fig:imgs2x}
    \vspace{-0.1in}
\end{figure}

 We chose $\lambda=2\times 10^{-4}$ and $\beta=1\times 10^{-4}$. We optimized for 30 iterations in total. To accelerate the convergence, we gradually increased $\beta$ to $4\times10^{-4}$ and $8\times10^{-4}$ in the last two iterations, respectively. To solve $\bs{\mc Z}^{n+1}$, we applied a 2-step steepest gradient descent with a step size of 0.1. The magnitude of the input slices was normalized to be less than one and recovered to the original scale afterward. The maximum number of steps of CG was 30 with an exit residual threshold of 0.05 to obtain $\mb U^{n+1}$. 
\vspace{-0.05in}
\section{Experiments and Results}
\vspace{-0.05in}
\subsection{Comparison with SOTA methods}
\vspace{-0.1in}

We compare the proposed method to the three state-of-the-art (SOTA) methods discussed below:  1. Deep Factor Model (DFM) \cite{chen2023deep} - a self-supervised method which employs Deep Image Prior (DIP) to reconstruct all contrasts of images jointly. 2.  A subspace method - this method imposes a quadratic regularization on the spatial factor i.e., $\|\mb U\|_2^2$ with a weight of $1\times 10^{-3}$.  
3. Subspace + wavelet method - this method uses $\|\mb W \mb U\|_1$ as the regularization, where $\mb W$ is a wavelet operator with weight of $2\times 10^{-4}$. All hyper-parameters were chosen to yield the best image quality as well as accurate $T_1$ estimates. 
In this work, we consider the reconstruction of a 4.5-min MPnRAGE scan, where the reference is recovered using DFM from a 9-min scan. The $T_1$ map was estimated from MPnRAGE using multi-pass fitting with spatial smoothing of $B_1$, inversion efficiency maps, and final $T_1$, as reported in \cite{kecskemeti2016mpnrage}. We performed one-pass fitting for the $T_1$ estimation of the SOTA and the proposed method. PSNR and error maps were calculated between the estimated and the reference $T_1$ map. 

The recovered multi-contrast images are shown in Fig. \ref{fig:imgs2x}. We note that the subspace and subspace + wavelet methods in Fig. \ref{fig:imgs2x} (b) \& (c) offer good quality reconstructions when the inversion time (TI) is high (e.g. TI = 900.16 ms). It should be noted that the magnetization changes slower during this period, which makes the reconstruction easier. On the contrary, it is challenging to recover images with shorter TI (e.g. 21.76 ms) due to the rapidly changing magnetization, which translates to a limited number of acquired projection angles with similar contrast. We observe that the DFM approach offers reduced artifacts at most TIs. However, the self-supervised training of DFM is computationally expensive (e.g., 8 hours for a 4.5-min scan, while SS-MuSE translates to a reduced run-time of 3 hours). We highlight that SS-MuSE achieves comparable denoising performance, while retaining more details, especially at TI = 21.76 and 900.16 ms as shown in Fig. \ref{fig:imgs2x}.

\begin{figure}[t!]
    \centering
    \includegraphics[width=0.5\textwidth]{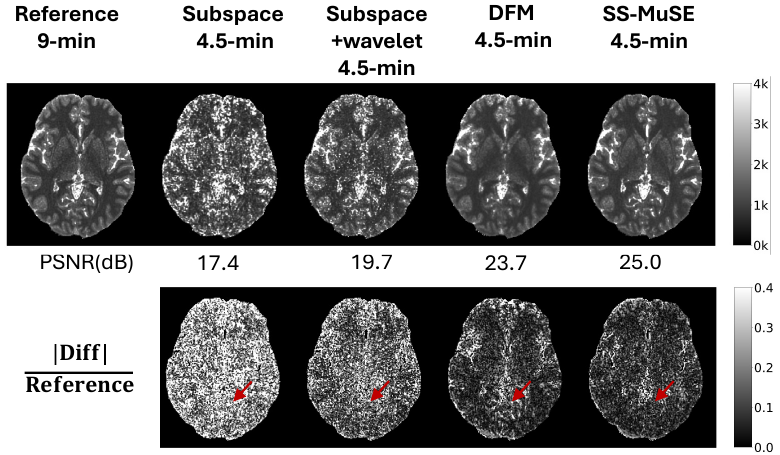}
    \caption{Comparison of $T_1$ maps: The Reference $T_1$ map is from the 9-min MPnRAGE scan. $|$Diff$|$ denotes $|$Reference - estimated $T_1$ map$|$. Subspace and Subspace + wavelet suffer from a high variance of $T_1$ estimation due to aliasing artifacts in source images. SS-MuSE offers less bias than DFM as evidenced by higher PSNR. }
    \label{fig:t1map}
    \vspace{-0.1in}
\end{figure}

The reconstructed T1 maps from the different methods are shown in Fig. \ref{fig:t1map}. T1 maps from subspace and subspace + wavelet appear highly noisy. The errors from DFM and SS-MUSE are significantly smaller on the contrary. We note that SS-MuSE offers slightly sharper and more accurate T1 maps than DFM, which can also appreciated from the higher PSNR compared to DFM.

\section{Conclusion}
\vspace{-0.05in}
We introduced SS-MuSE to enable fast iterative reconstruction of multicontrast MRI data using a multiscale energy (MuSE) model as a denoiser. SS-MuSE achieves better image quality than unsupervised methods while offering faster computation time than self-supervised methods. 
SS-MuSE offers improved source images and $T_1$ maps, compared to existing methods. 
\vspace{-3mm}\section{Compliance with ethical standards}\vspace{-3mm}
The data was acquired on a 3T GE Premier MRI scanner at the University of Iowa (UI). This study was approved by the institutional review board (IRB) at UI, and informed written consent was obtained prior to scanning. 

\vfill\pagebreak
\bibliographystyle{IEEEbib}
\bibliography{refs}

\begin{thebibliography}{10}

\bibitem{kecskemeti2016mpnrage}
Steven Kecskemeti, Alexey Samsonov, Samuel~A Hurley, Douglas~C Dean, Aaron Field, and Andrew~L Alexander,
\newblock ``Mpnrage: A technique to simultaneously acquire hundreds of differently contrasted mprage images with applications to quantitative t1 mapping,''
\newblock {\em Magnetic resonance in medicine}, vol. 75, no. 3, pp. 1040--1053, 2016.

\bibitem{mugler1990three}
John~P Mugler~III and James~R Brookeman,
\newblock ``Three-dimensional magnetization-prepared rapid gradient-echo imaging (3d mp rage),''
\newblock {\em Magnetic resonance in medicine}, vol. 15, no. 1, pp. 152--157, 1990.

\bibitem{ravishankarLowRankAdaptiveSparse2017}
Saiprasad Ravishankar, Brian~E. Moore, Raj~Rao Nadakuditi, and Jeffrey~A. Fessler,
\newblock ``Low-{{Rank}} and {{Adaptive Sparse Signal}} ({{LASSI}}) {{Models}} for {{Highly Accelerated Dynamic Imaging}},''
\newblock vol. 36, no. 5, pp. 1116--1128.

\bibitem{zhaoAcceleratedMRParameter2015}
Bo~Zhao, Wenmiao Lu, T.~Kevin Hitchens, Fan Lam, Chien Ho, and Zhi-Pei Liang,
\newblock ``Accelerated {{MR}} parameter mapping with low-rank and sparsity constraints,''
\newblock vol. 74, no. 2, pp. 489--498.

\bibitem{qiuDirectSynthesisMulticontrast2023}
Shihan Qiu, Sen Ma, Lixia Wang, Yuhua Chen, Zhaoyang Fan, Franklin~G. Moser, Marcel Maya, Pascal Sati, Nancy~L. Sicotte, Anthony~G. Christodoulou, Yibin Xie, and Debiao Li,
\newblock ``Direct synthesis of multi-contrast brain {{MR}} images from {{MR}} multitasking spatial factors using deep learning,''
\newblock vol. 90, no. 4, pp. 1672--1681.

\bibitem{aggarwal2018modl}
Hemant~K Aggarwal, Merry~P Mani, and Mathews Jacob,
\newblock ``Modl: Model-based deep learning architecture for inverse problems,''
\newblock {\em IEEE transactions on medical imaging}, vol. 38, no. 2, pp. 394--405, 2018.

\bibitem{sandino2021deep}
Christopher~Michael Sandino, Frank Ong, Siddharth~Srinivasan Iyer, Adam Bush, and Shreyas Vasanawala,
\newblock ``Deep subspace learning for efficient reconstruction of spatiotemporal imaging data,''
\newblock in {\em NeurIPS 2021 Workshop on Deep Learning and Inverse Problems}, 2021.

\bibitem{chen2022data}
Zihao Chen, Yuhua Chen, Yibin Xie, Debiao Li, and Anthony~G Christodoulou,
\newblock ``Data-consistent non-cartesian deep subspace learning for efficient dynamic mr image reconstruction,''
\newblock in {\em 2022 IEEE 19th International Symposium on Biomedical Imaging (ISBI)}. IEEE, 2022, pp. 1--5.

\bibitem{jun2024zero}
Yohan Jun, Yamin Arefeen, Jaejin Cho, Shohei Fujita, Xiaoqing Wang, P~Ellen Grant, Borjan Gagoski, Camilo Jaimes, Michael~S Gee, and Berkin Bilgic,
\newblock ``Zero-deepsub: Zero-shot deep subspace reconstruction for rapid multiparametric quantitative mri using 3d-qalas,''
\newblock {\em Magnetic Resonance in Medicine}, vol. 91, no. 6, pp. 2459--2482, 2024.

\bibitem{liuMRParameterMapping2021}
Fang Liu, Richard Kijowski, Georges El~Fakhri, and Li~Feng,
\newblock ``{{MR Parameter Mapping Using Model-Guided Self-Supervised Deep Learning}},''
\newblock vol. 85, no. 6, pp. 3211--3226.

\bibitem{chen2023deep}
Yan Chen, James~H Holmes, Curtis Corum, Vincent Magnotta, and Mathews Jacob,
\newblock ``Deep factor model: A novel approach for motion compensated multi-dimensional mri,''
\newblock in {\em 2023 IEEE 20th International Symposium on Biomedical Imaging (ISBI)}. IEEE, 2023, pp. 1--4.

\bibitem{chand2024multi}
Jyothi~Rikhab Chand and Mathews Jacob,
\newblock ``Multi-scale energy (muse) framework for inverse problems in imaging,''
\newblock {\em IEEE transactions on computational imaging}, 2024.

\bibitem{venkatakrishnan2013plug}
Singanallur~V Venkatakrishnan, Charles~A Bouman, and Brendt Wohlberg,
\newblock ``Plug-and-play priors for model based reconstruction,''
\newblock in {\em 2013 IEEE global conference on signal and information processing}. IEEE, 2013, pp. 945--948.

\bibitem{ahmad2020plug}
Rizwan Ahmad, Charles~A Bouman, Gregery~T Buzzard, Stanley Chan, Sizhuo Liu, Edward~T Reehorst, and Philip Schniter,
\newblock ``Plug-and-play methods for magnetic resonance imaging: Using denoisers for image recovery,''
\newblock {\em IEEE signal processing magazine}, vol. 37, no. 1, pp. 105--116, 2020.

\bibitem{lee2023improving}
Suhyeon Lee, Hyungjin Chung, Minyoung Park, Jonghyuk Park, Wi-Sun Ryu, and Jong~Chul Ye,
\newblock ``Improving 3d imaging with pre-trained perpendicular 2d diffusion models,''
\newblock in {\em Proceedings of the IEEE/CVF International Conference on Computer Vision}, 2023, pp. 10710--10720.

\end{thebibliography}

\end{document}